# Two-Dimensional Boron Oxides with Dirac Loop and Strongly anisotropic Carrier Mobility

Ruiqi Zhang,[a] and Jinlong Yang[a,b] *

[a] Hefei National Laboratory for Physical Sciences at the Microscale, University of Science and Technology of China, Hefei, Anhui 230026, China.

[b] *Synergetic Innovation Center of Quantum Information & Quantum Physics, University of Science and Technology of China, Hefei, Anhui 230026, China.*

KEYWORDS. First-principles calculations, 2D boron oxides, Dirac loop, anisotropic carrier mobility

**ABSTRACT:** Recently, two-dimensional boron sheets have attracted a lot of attentions owing to their structural polymorphs and outstanding properties. And, due to chemical complexity and electron deficiency of B atoms, the 2D boron sheets are easy affected by the environment. So, exploring novel 2D boron oxides gets highly needed. In this study, we theoretically explored 2D boron oxides structures and their basic properties. We found 2D boron oxides are metals or semimetals, when oxygen concentration is low. More interesting, the $B_6O_1$ exhibits Dirac Loop near the Fermi level and the maximum Fermi velocity is estimated as high as $0.61 \times 10^6$ m/s, which much close to that in graphene. In addition, when the oxygen concentration is one forth, the most stable $B_4O_1$ get a semiconductor with a direct band gap of 1.24 eV and a strong anisotropic carrier mobility. Such huge differences of carrier mobility between two directions have not been reported in other 2D materials. Besides, when oxygen concentration is higher than one forth, the 2D boron sheets will get unstable based our results. We also proposed a method how to synthesize these systems. Our work provide a basic knowledge of 2D boron oxides and may lead to novel 2D boron oxides discovered. The unique electronic properties of the $B_6O_1$ and $B_4O_1$ render them promising 2D materials for applications in high-performance nanodevices.

## INTRODUCTION

Since the successful isolation of graphene in 2004,[1] the synthesis of novel two-dimensional (2D) materials have attracted tremendous interest due to their fascinating electronic, mechanical, optical or thermal properties.[2–5] As a typical example, graphene is famous for its linear Dirac-like dispersion near the Fermi level, in which many unusual electronic and spintronic properties were discovered.[3] However, the absence of a fundamental band gap severely limits its applications in field-effect transistors (FETs). Subsequently, possessing intrinsic direct band gaps and a considerable mobility, monolayer transition-metal dichalcogenides (TMDs) have attracted wide interest.[6–8] Nevertheless, since TMDs are compounds, the stoichiometry of 2D TMD is particularly challenging to control during growth by chemical vapor deposition (CVD) or in patterning lithographies.[9–11] Recently, with a sizable and novel direct band gap, high mobility, anisotropic electric conductance and optical response and so on, phosphorene is considered to be superior to graphene in electronic and optoelectronic applications.[12–16] However, phosphorene has a poor stability, because it is easy degradation when exposed to air.[17–21] Thus, developing 2D materials with high stability, fascinating electronic and optical properties is highly demanded.

Most recently, monolayer 2D boron sheets have been reported in the experiment.[22,23] Owing to their fascinating physical and chemical properties, such as considerable toughness, highly anisotropic metallic, superconducting and so on, they are considered to have the potential to be an important 2D material in future nanodevices.[22,24,25] However, due to the chemical complexity and electron deficiency of B atoms, extraordinary 2D boron sheets were discovered in experiment, such as borophene, $\beta_{12}$ and $\chi_3$ sheet, and monolayer $\gamma$-$B_{28}$.[22,23,26] Besides, various forms of monolayer boron structures with outstanding properties have been predicted by theoretical work.[27–32] In addition, some study have reported that monolayer boron structures can be mediated by metal substrates.[23,26,29] All these indicate that it is easy to be affected/oxidized by the environments. It is notable that B-O bond peak have been observed when 2D boron sheets exposed to in Feng's work.[23] And they found the boron sheet is metallic while the oxidized boron is insulating. What is more, we also note that Tai et al. use a mixture of boron and boron oxide powders as the boron source to grow boron thin film by CVD method.[26] So these signs suggest that the 2D boron oxides may be exist. On the other hand, like graphene and phosphorene oxides,[17,33–35] 2D boron oxides may be have fascinating properties and can be applied in future nanodivces. Hence, it would be interesting to consider how the 2D boron sheet affected by oxidation and 2D boron oxides exist in what form? Thus, a detailed study of the structure of 2D boron oxides is highly needed, which not only give us a basic knowledge of 2D boron oxides but also may lead to discover new functional 2D materials.

Herein, in this work, we systematically performed a global search[35–37] to predict the lowest energy structures of 2D boron oxides with different oxygen concentrations by using particle-swarm optimization (PSO) algorithm[38] combined with first-principles methods. The oxygen concentration changed from 0.125 to 0.500. We found that 2D boron oxides show metallic and semi-metallic when the oxygen concentration is low. While, when the oxygen concentration is one forth, the 2D boron oxides will become semiconductor. However, when the oxygen concentration is higher than one forth, the 2D boron oxides will become unstable based on our calculation, which shown in Figure S1. Remarkably, the most stable structure of 2D $B_6O_1$ exhibits Dirac Loop near the Fermi level with a high Fermi velocity which is close to that in graphene.[39] For the most stable 2D $B_4O_1$, it has a direct band gap of 1.24 eV and shows highly anisotropic carrier mobility. In one direction, the electrons has a high mobility up to 267.451 $cm^2/V/s$, while in the other one, that is very slow with a mobility of $8.383 \times 10^{-4}$ $cm^2/V/s$. To our knowledge, such huge anisotropy have not been reported in other 2D materials. These outstanding and unique characters will make these 2D boron oxides very attractive for application in nano-electronic and optoelectroic devices. Herein, at first we discuss the basic characters of 2D boron oxides. Because, the $B_8O_1$, $B_7O_1$ and $B_5O_1$ show metallic (see Figure S2) and the 2D boron oxides with oxygen concentration are not stable. So, we then focus on 2D $B_6O_1$ and $B_4O_1$ to show that their unique electronic properties and they are promising materials for using in nanodivces.

## COMPUTATIONAL DETAILS

In this study, our first-principles calculations are based on the density functional theory[40] (DFT) and implemented in the Vienna ab initio simulation package (VASP) package.[41,42] The atomic structure of every $B_xO_y$ is constructed in the xy-plane. A large value ~20 Å of the vacuum region along the z-direction is used to avoid interaction between two adjacent periodic images. The generalized gradient approximation (GGA) of Perdew, Burke, and Ernzerhof [43] (PBE) functional and projector augmented wave[44,45] (PAW) potentials are used. In all computations, the kinetic energy cutoff are set to be 520 eV in the plane-wave expansion. All the geometry structures are fully relaxed until energy and forces are converged to $10^{-5}$ eV and 0.01 eV/Å, respectively. Moreover, accurate hybrid functional calculations using HSE06 [46,47] was used to calculate more reliable band structure of 2D P4O1. Phonon dispersion analysis was performed using the PHONOPY[48] code interfaced with the density functional perturbation theory (DFPT)[49] as implemented in VASP. The ab initio molecular dynamics (AIMD) simulations are also performed with VASP. Each MD simulation in NVT ensemble lasted for 16 ps with a time step of 2.0 fs.

The carrier mobility can be calculated using the deformation potential (DP) model[50] based on the effective mass approximation, The mobility for the 2D system is expressed as[51,52]

$$\mu_{2D} = \frac{2e\hbar^3 C_{2D}}{3k_B T (m^*)^2 E_1^2} \quad (1)$$

Where T is the temperature, and $m^* = \hbar^2 \left[\frac{\partial^2 E(k)}{\partial k^2}\right]^{-1}$ is the effective mass of the carrier along the transport direction. $E_1 = \Delta E/(\Delta l/l_0)$, $\Delta E$ is the energy of the band edges change, and $l_0$ is the lattice constant in the transport direction, and $\Delta l$ is its deformation. C is the elastic modulus defined as $C_{2D} = [\partial^2 E/\partial \delta^2]/S_0$, where E is the total energy of supercell, and δ stands for the applied strain, and $S_0$ is the area of cell.

Extensive PSO simulations are performed to find out the global stable structures of 2D B8O1, B7O1, B6O1, B5O1, B4O1, B3O1 and B2O1. We set the population size to 30 and the number of generations to 20. The total number of atoms in the unit cell is less than 14. Moreover, to make our results reliable, we repeat tiwce of ecah calculation.

## RESULTS AND DISCUSSIONS

From our PSO simulations, we found the stable monolayer $B_xO_y$ sheet displayed different motifs when the oxygen concentration varied. Among of them, the B8O1, B7O1 and B5O1 can be viewed as triangular lattices and hexagonal holes. In B8O1 nanosheet, both B-B and B-O hexagonal holes are exits. While, in the B7O1 and B5O1, there are only B-O hexagonal holes. Unlike the others, the B6O1 is a sandwich structure where the B triangular lattices in the outside layer and the O atoms in the interlayer. For the B4O1, except the triangular lattices and B-O hexagonal holes, there also exits quadrilateral lattices consisting of B and O atoms and all the O atoms are surrounded by three B atoms. All these 2D $B_xO_y$ sheet are non-planar structures. Among of them, the B6O1 has the largest thickness of 4.592 Å and the B5O1 have the smallest thickness of 0.137 Å. All these results suggest the B and O can form diverse structures. All the optimized layered structures are presented in Figure 1.

To access the experimental feasibility of the predicted 2D boron oxides, we first calculated the average cohesive energy, which is defined as $E_{coh} = (xE_B + yE_O - E_{B_xO_y})/(x+y)$, where $E_B$, $E_O$, and $E_{B_xO_y}$ are the total energies of a single B atom, a single O atom, and one $B_xO_y$ unit cell, respectively. Remarkably, the calculated results show that the cohesive energy for these monolayer are close to that of synthesized borophene (6.242 eV/atom) and $β_{12}$ sheet (6.282 eV/atom)[32] and much higher silicene (3.71 eV/atom) and Cu2Si (3.46 eV/atom) at the same calculation level.[52–54] High cohesive energy of these structures clearly demonstrates that the bonding in 2D $B_xO_y$ sheet is robust and they have high possibility for experimental realization. Table 1 lists the lattice constants and energies for these 2D boron oxides.

**Table 1. The Calculated the Lattice Constant, Symmetry Group (SG) and Cohesive Energy ($E_C$) for the 2D boron oxides sheet.**

| Phase | a | b | c | SG | $E_{coh}$ |
|---|---|---|---|---|---|
| B8O1 | 9.427 | 2.888 | 20.000 | P2/M | 6.053 |
| B7O1 | 7.732 | 2.883 | 20.000 | PMM2 | 6.034 |
| B6O1 | 2.884 | 2.884 | 20.000 | P6/MMM | 6.124 |
| B5O1 | 6.168 | 2.907 | 20.000 | PMM2 | 6.035 |
| B4O1 | 9.218 | 2.799 | 20.000 | P21/M | 6.145 |

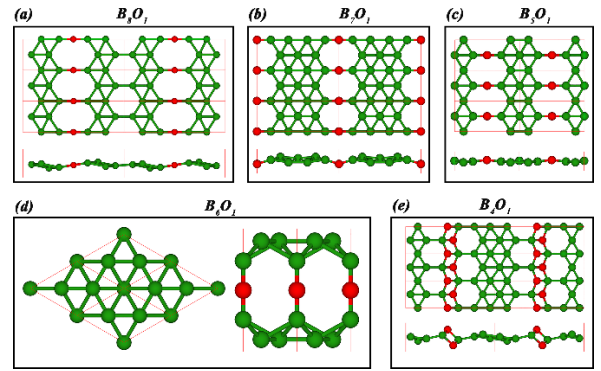

**Figure 1. Top and Side views of four the lowest-energy structures of (a) B8O1, (b) B7O1, (c) B5O1, (d) B6O1 and (e) B4O1. The green and red balls represent B atoms and O atoms, respectively. The shapes of unit cell are labeled with red solid line.**

Electronic Properties

To obtain a thorough knowledge of these 2D $B_xO_y$, we then studied the electronic properties of these compounds. Based on our calculations, the $B_8O_1$, $B_7O_1$ and $B_5O_1$ show metallic and their band structures are illustrated Figure S1. However, the $B_6O_1$ show semimetal with a Dirac loop and the $B_4O_1$ is a semiconductor with strongly anisotropic carrier mobility. Then we put our attentions on the study of $B_6O_1$ and $B_4O_1$.

$B_6O_1$: The band structures and density of states for $B_6O_1$ are depicted in Figure 2. As showed in Figure 2 (a), the valence and conduction bands meet near the Fermi level and show the linear dispersion relation. There are two cross points located at asymmetric positons: one is at (0.305, 0.305, 0.000) along Γ to K and another one is at (0.343, 0.274, 0.000) along K to M. As plotted in Figure 2(b), the DOS of $B_6O_1$ sheet is zero at the Fermi level and the DOS near the Fermi level are contributed by B atoms. Be different from conventional Dirac 2D materials, just like graphene,[3] γ-borophene[28] and 2D P6/mmm boron[32], the $B_6O_1$ form a Dirac loop or not Dirac cone near the Fermi Level, which is shown in Figure 2 (c). Then, we estimated the Fermi velocity according to $v_f \approx \partial E/\partial k$. The outer (iner) radial velocities of the states on the loop have the largest value of $0.61 \times 10^6$ m/s ($0.50 \times 10^6$ m/s) along the high symmetry line Γ–K and the lowest value of $0.26 \times 10^6$ m/s ($0.22 \times 10^6$ m/s) along the line K-M. Thus the Fermi velocity in 2D B6O1 are oscillated along the node-loop in the range from $0.26 \times 10^6$ m/s to $0.61 \times 10^6$ m/s.

To further explore the origin of Dirac loop, the PDOS and the band-decomposed charge density at Dirac loop are plotted in Figure S3. Figure S3 (b) show that the Dirac loop are mainly derived from $p_z$ orbitals of B atoms. And, the hybrids $p_z$ orbitals from covalent B atoms form π bands and $π^*$ bands. These bands meet and show linear dispersion relation near the Fermi surface. So the origin of Dirac loop of the $B_6O_1$ can be explained by the crossing of π and $π^*$ bands come from the $p_z$ orbitals of B atoms, which is the same with graphene[3] and T-graphene.[55]

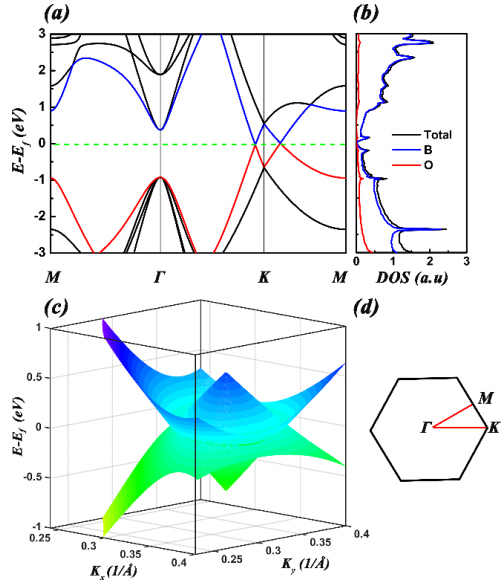

**Figure 2. The calculated electronic structure of the 2D B6O1. (a) Band structures, (b) DOS and (c) Dirac loop formed by the valence and conduction bands, (d) Γ (0.0, 0.0, 0.0), M (0.5, 0.5, 0.0), and K (1/3, 1/3, 0.0) refer to special points in the first Brillouin zone.**

$B_4O_1$: The computed band structure and DOS of monolayer $B_4O_1$ sheet using HSE06 is shown in Figure 3 (a). Obviously, it is a semiconductor with a direct band gap of 1. 24 eV. The positions of CBM and VBM are located at (0.00, 0.412, 0.000). In addition, the band structures near the CBM and VBM of monolayer $B_4O_1$ exhibit notable dispersion behavior from X (0.500, 0.500, 0.00) to S (0.500, 0.500, 0.000) and Y (0.000, 0.500, 0.000) to Γ (0.000, 0.000, 0.000)), indicating that the 2D B4O1 monolayer may have relatively high carrier mobility along Y direction. Surprisingly, the band structures are almost a straight line from S (0.500, 0.500, 0.000) to Y (0.000, 0.500, 0.000), suggesting that $B_4O_1$ may have an ultralow carrier mobility along X direction. According to eq.1, to predict the carrier mobility of monolayer $B_4O_1$, we first calculated DP Constant ($E_1$), 2D Elastic Modulus ($C_{2D}$) and Effective Mass. As show in Table 2, the 2D modulus along X is a little bigger than along y direction. It can be explained by that the chemical bonds strength are different along different directions. Because the length of B-B bond along X direction is almost the same as that along Y direction, it may be the B-O bond strength play a crucial role. Actually, the length of B-O bond is 1.531 Å along X direction and that is 1.605 Å along Y direction. In addition, the effective mass shows an obviously anisotropic feature. For electrons, $m_e^*$ is 767.119 $m_e$ along X direction, which is almost 700 times greater than that (1.103 $m_e$) in the y direction. While for holes, the effective mass $m_h^*$, are 565.916 $m_e$ and 1.191 $m_e$ in the X and Y direction, respectively. These results can be well explained by the charge-density plot of CBM and VBM in Figure 3 (b) and (c), where both CBM and VBM are localized in the X direction, while are delocalized in the Y direction.

As a result of the strongly anisotropic of effective mass, the carrier mobility is highly anisotropic along different direction. In the Y direction, the electron mobility is 267.451 cm$^2$/V/s, which is about $3.2\times10^5$ times larger than that in X direction. And, the hole mobility is 90.496 cm$^2$/V/s in Y direction which is about $3.5\times10^4$ times bigger than that in X direction. As a result of these, the electron and hole will almost transport along Y direction and not move along X direction. To our knowledge, these strongly anisotropic for electron and hole mobility along different direction is unique among prevailing 2D materials. Such a unique property can be make 2D $B_4O_1$ have a unique application in nanodivces, such as we can use 2D $B_4O_1$ to do a special logic circuits. It should note that the predicted electron (hole) mobility of ∼267.451 (∼90.496) cm$^2$/V/s in Y direction, which have a quite similar levels of 2D MoS$_2$ and monolayer phosporene.[15,51]

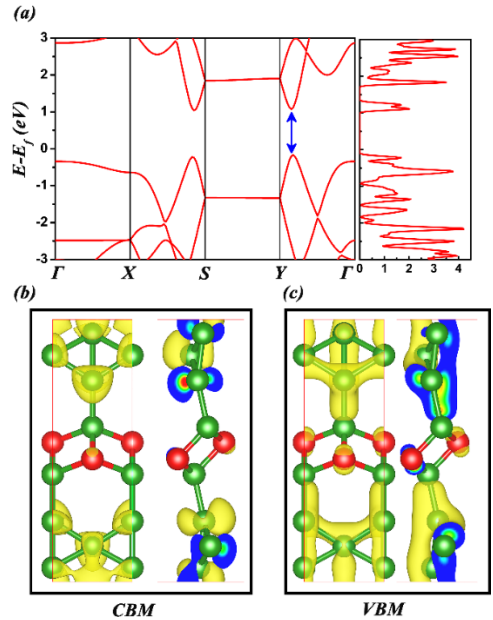

**Figure 3. The calculated band structure (a) and DOS (b) of the 2D B$_4$O$_1$ by using HSE06 function. (c) and (d) are the charge density corresponding to the CBM and VBM for the monolayer B$_4$O$_1$, respectively. The isovalue is 0.008 e/Bohr$^3$**

**Table 2. The Calculated DP Constant ($E_1$), 2D Elastic Modulus ($C_{2D}$), Effective Mass (m*), and Electron and Hole Mobility (μ) along x and y Direction of the B4O1 Sheet at 300 K**

| carrier (direction) | $E_1$ (eV) | $C_{2D}$(J/m$^2$) | m*(m$_e$) | μ(cm$^2$/V/s) |
|---|---|---|---|---|
| e (x) | 1.586 | 87.393 | 767.119 | 8.383× 10$^{-4}$ |
| h (x) | 1.222 | 87.393 | 565.916 | 2.595× 10$^{-3}$ |
| e (y) | 1.797 | 74.003 | 1.103 | 267.451 |
| h (y) | 2.861 | 74.003 | 1.191 | 90.496 |

*Stability and Experimental Synthesis*: At last, to test the kinetic stability of monolayer $B_xO_y$ sheet, we computed their phonon spectrum. The phonon dispersions are shown in Figure 4 and Figure S4. Obviously, the absence of any imaginary phonon modes demonstrate that these structures are dynamically stable. Especially, the highest frequency of monolayer $B_xO_y$ reach up to about 1400 cm$^{-1}$, which are much higher than the highest frequency of 459 cm$^{-1}$ in phosphorene[56] and 473 cm$^{-1}$ in MoS$_2$ monolayer[57], indicating robust B-B and B-O bonds in these monolayer sheet. In addition, the thermal stability is another important indicator to test structures stability. So, AIMD is performed at DFT level for selected systems. The NVT ensemble is used. In this simulation, all the initial structures are used 3 × 3 × 1 supercell and the temperature of BOMD simulations are controlled at 300K. Snapshots at 16 ps for each monolayer $B_xO_y$ were plotted in Figure S5. Only the structure of 2D B4O1 break, suggesting that it may not thermally stable at room temperature. The other structural decompositions were not observed suggesting these structures are thermal stability. Note that AIMD simulation here is artificial due to the limitation of simulation time scale and size of supercells.

Our results confirmed these structures are kinetic and thermal stability. So the question for how to synthesize these 2D sheet will be get more important to us. It is note that although many borophene allotrope were observed in experiment, there are new borophene allotrope were discovered using different methods and metal substrates. One method of them is using a mixture of boron and boron oxide powders as the boron source to grow boron thin film and γ-B$_{28}$ monolayer was found.[26] So if we want prepared the 2D Boron Oxides, we can use different proportion of boron and boron oxide as the source and selected suitable gas the carrier gas to prepared different 2D $B_xO_y$ . If monolayer boron oxide were synthesized in experiment, it will greatly broaden the application of borphene.

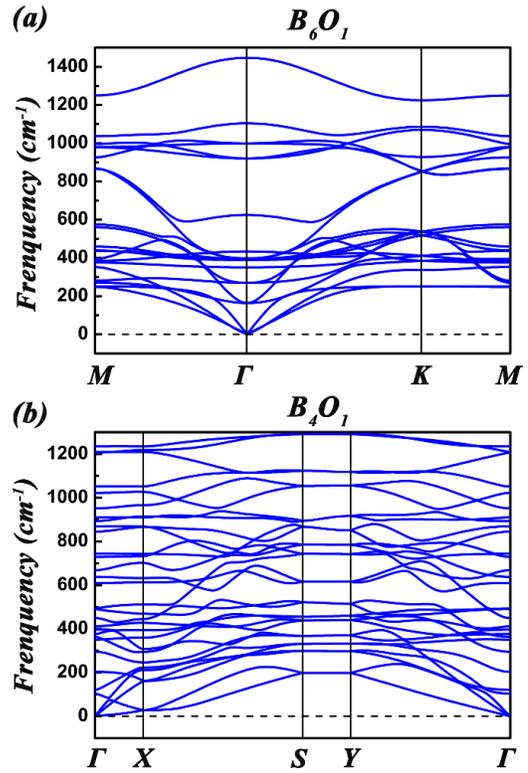

**Figure 4.** The calculated phonon band structures of the 2D (a) B6O1 and (b) B5O1.

CONCLUSION

In conclusion, by global optimization method we systematically searched the most stable structures of 2D B$_8$O$_1$, B$_7$O$_1$, B$_6$O$_1$, B$_5$O$_1$, B$_4$O$_1$, B$_3$O$_1$ and B$_2$O$_1$. Based on our calculations, the lowest B$_3$O$_1$ and B$_2$O$_1$ are not stable, which suggest they may can't exists. The B$_8$O$_1$, B$_7$O$_1$ and B$_5$O$_1$ have nonplanar structures and shows metallic. While, the lowest energy of the B$_6$O$_1$ is a sandwich structures. Most strikingly, this structures exhibits Dirac loop near the Fermi level and have a Fermi velocity as high as 0.61 × 10$^6$ m/s. In addition, the B$_4$O$_1$ is demonstrated to be a semiconductor with a moderate direct band gap of 1.24 eV. At the same time, the electron and hole mobility of the B$_4$O$_1$ sheet is highly anisotropic. More specifically, the electron mobility in X direction exhibits a high value of 267.451 cm$^2$/V/s, which is larger than that in Y direction by 3.2×10$^5$ times. These hugely anisotropic carrier mobility have not been reported in other 2D materials. The kinetic and dynamically stable suggest they may can be synthesized. We also proposed a method to synthesize these 2D sheet. Our findings show that 2D boron oxides can be server as novel 2D materials with unique electronic properties and shed new light on the electronic and structure diversity of 2D boron oxides.

## ASSOCIATED CONTENT

**Supporting Information**. Optimized structures of 2D B3O1, B2O1-1and B2O1-2 and their corresponding phonon band structures. The calculated band structures of the 2D B$_8$O$_1$, B$_7$O$_1$ and B$_5$O$_1$ by using PBE function. The charge density corresponding to the CBM and VBM for the monolayer B$_6$O$_1$. And, the PDOS of 2D B$_6$O$_1$. The calculated phonon band structures of the 2D B8O1, B7O1 and B5O1. And snapshots of AIMD simulations are provided.

## AUTHOR INFORMATION

**Corresponding Author**

* Corresponding Author Email: jlyang@ustc.edu.cn
**Notes**
The authors declare no competing financial interest.

## ACKNOWLEDGMENT
This work is partially supported the National Key Basic Research Program (2011CB921404), the NSFC (21421063, 91021004, 21233007), the Chinese Academy of Sciences (CAS) (XDB01020300), and by USTCSCC, SCCAS, Tianjin, and Shanghai Supercomputer Centers.